\documentclass[10pt,  letterpaper]{IEEEtran}
\usepackage{todonotes}
\usepackage[nocompress]{cite}
\usepackage{url}
\usepackage{enumitem}
\usepackage{epstopdf}
\usepackage{graphicx}
\usepackage{caption}
\usepackage{listings}
\usepackage{amsmath}
\usepackage{tabu}
\usepackage{multirow}
\usepackage{adjustbox}
\usepackage{fontawesome}
\usepackage{algpseudocode}
\usepackage{soul}
\usepackage[bookmarks=false]{hyperref}
\usepackage{footnote}
\usepackage{lipsum}
\usepackage[ruled,vlined, linesnumbered]{algorithm2e}
\usepackage{subcaption}
\usepackage{booktabs}
\usepackage{tikz}
\usetikzlibrary{positioning}
\soulregister\cite7
\soulregister\ref7
\soulregister\pageref7
\soulregister\secref7
\soulregister\tabref7

\newcommand*\circledw[1]{\tikz[baseline=(char.base)]{
            \node[shape=circle,draw,inner sep=0.75pt, text=black,fill=white] (char) {#1};}}

\newcommand{\shrinkspace}{\vspace{-5mm}}

\definecolor{ao}{rgb}{0.0, 0.5, 0.0}
\definecolor{db}{rgb}{0.2, 0.2, 0.6}
\definecolor{cadmiumgreen}{rgb}{0.0, 0.42, 0.24}

\newcommand{\tabref}[1]{Table~\ref{#1}}

\newcommand{\secref}[1]{Section~\ref{#1}}

\SetAlFnt{\footnotesize}

\begin{document}
\title{gFaaS: Enabling Generic Functions in Serverless Computing}


\author{\IEEEauthorblockN{Mohak Chadha, Paul Wieland, Michael Gerndt\\}
\IEEEauthorblockA{Chair of Computer Architecture and Parallel Systems, Technische Universit{\"a}t M{\"u}nchen \\
Garching (near Munich), Germany\\}
Email: \{firstname.lastname\}@tum.de}

\maketitle
\vspace{-20mm}


\begin{abstract}
%
With the advent of AWS Lambda in 2014, Serverless Computing, particularly Function-as-a-Service (FaaS), has witnessed growing popularity across various application domains. FaaS enables an application to be decomposed into fine-grained \emph{functions} that are executed on a FaaS platform. It offers several advantages such as no infrastructure management, a pay-per-use billing policy, and on-demand fine-grained autoscaling. However, despite its advantages, developers today encounter various challenges while adopting FaaS solutions that reduce productivity. These include FaaS platform lock-in, support for diverse function deployment parameters, and diverse interfaces for interacting with FaaS platforms. To address these challenges, we present \emph{gFaaS}, a novel framework that facilitates the holistic development and management of functions across diverse FaaS platforms. Our framework enables the development of \emph{generic} functions in multiple programming languages that can be seamlessly deployed across different platforms without modifications. Results from our experiments demonstrate that \emph{gFaaS} functions perform similarly to native platform-specific functions across various scenarios. A video demonstrating the functioning of \emph{gFaaS} is available from \url{https://youtu.be/STbb6ykJFf0}.
\end{abstract}






\lstdefinestyle{htmlcssjs} {%
  basicstyle={\footnotesize\ttfamily},   
  frame=b,
  xleftmargin={0.75cm},
  numbers=left,
  stepnumber=1,
  firstnumber=1,
  numberfirstline=true,	
  identifierstyle=\color{black},
  keywordstyle=\color{blue}\bfseries,
  ndkeywordstyle=\color{editorGreen}\bfseries,
  stringstyle=\color{editorOcher}\ttfamily,
  commentstyle=\color{brown}\ttfamily,
  language=HTML5,
  alsolanguage=JavaScript,
  alsodigit={.:;},	
  tabsize=2,
  showtabs=false,
  showspaces=false,
  showstringspaces=false,
  extendedchars=true,
  breaklines=true,
  literate=%
  {Ö}{{\"O}}1
  {Ä}{{\"A}}1
  {Ü}{{\"U}}1
  {ß}{{\ss}}1
  {ü}{{\"u}}1
  {ä}{{\"a}}1
  {ö}{{\"o}}1
}

\lstset{language=java,
                keywordstyle=\color{blue}\ttfamily,
                stringstyle=\color{red}\ttfamily,
                commentstyle=\color{ao}\ttfamily,
                numbers=left,
                numberstyle=\tiny,
                morecomment=[l][\color{magenta}]{\#},
                numbers=left,
                framexleftmargin=2mm,
                showstringspaces=false,
                numbersep=1mm,
                morekeywords={import, public, class, void, new, extends}
}

\section{Introduction}
\label{sec:intro}
\vspace{-1mm}


Serverless computing aka Function-as-a-Service (FaaS) is an emerging cloud computing paradigm~\cite{serverlessrise}, that has gained significant popularity and adoption in various application domains, such as machine learning~\cite{fedless, fedlesscan, serverlessfl, fastgshare},  edge computing~\cite{fado}, heterogeneous computing~\cite{fncapacitor, jindal2021function, courier, postericdcs}, and scientific computing~\cite{chadha2021architecture, demystifying}. In FaaS, developers implement fine-grained pieces of code called \emph{functions} that are packaged independently in containers and deployed onto a FaaS platform. Most FaaS platforms offer developers the flexibility to implement functions using a wide array of programming languages such as \texttt{Java}, \texttt{Node.js}, and \texttt{Go}.
FaaS functions are \textit{ephemeral}, i.e., short-lived, and \textit{event-driven}, i.e., these functions only get executed in response to external triggers such as HTTP or \texttt{gRPC}~\cite{grpc2016} requests. In addition, these functions are \textit{stateless}, i.e., any application state needs to be persisted in external storage or databases. Several commercial and open-source FaaS platforms, such as AWS Lambda~\cite{aws_lambda}, Google Cloud Functions (GCF)~\cite{google_cloud_functions}, OpenFaaS~\cite{openfaas}, Knative~\cite{knative_serving}, Nuclio~\cite{nuclio}, and Fission~\cite{fission} are currently available.

\begin{figure*}[t]
\begin{minipage}[t]{.48\textwidth}
\begin{lstlisting}[language=java, frame=single,  caption=OpenFaaS~\cite{openfaas_java_template}.,  captionpos=b, basicstyle=\ttfamily\tiny, belowskip=-1\baselineskip, label={fig:openfaasexample}]
package com.openfaas.function;
import com.openfaas.model.IHandler;
import com.openfaas.model.IResponse;
import com.openfaas.model.IRequest;
import com.openfaas.model.Response;

public class Handler extends com.openfaas.model.AbstractHandler {
    public IResponse Handle(IRequest req) {
        Response res = new Response();
	    res.setBody("Hello world!");
	    return res;
    }
}
\end{lstlisting}
\end{minipage}  
\hfill
\begin{minipage}[t]{.48\textwidth}
\begin{lstlisting}[language=java, frame=single,  caption=Fission~\cite{fission_java_model_function}.,  captionpos=b, basicstyle=\ttfamily\tiny, belowskip=-1\baselineskip, label={fig:fissionexample}]
package io.fission;
import org.springframework.http.RequestEntity;
import org.springframework.http.ResponseEntity;
import io.fission.Function;
import io.fission.Context;

public class HelloWorld implements Function {

  @Override
  public ResponseEntity<?> call(RequestEntity req, Context context) {
    return ResponseEntity.ok("Hello World!");
  }
}
\end{lstlisting}
\end{minipage}  
\hfill
\begin{minipage}[t]{0.48\textwidth}
\begin{lstlisting}[language=java, frame=single, caption=Nuclio~\cite{nuclio_java_model_reference}.,  captionpos=b,  basicstyle=\ttfamily\tiny, belowskip=-1\baselineskip, label={fig:nuclioexample}]
import io.nuclio.Context;
import io.nuclio.Event;
import io.nuclio.EventHandler;
import io.nuclio.Response;

public class EmptyHandler implements EventHandler {
    @Override
    public Response handleEvent(Context context, Event event) {
       return new Response().setBody("Hello world!");
    }
}
\end{lstlisting}
\end{minipage}  
\hfill
\begin{minipage}[t]{0.48\textwidth}
\begin{lstlisting}[language=java, frame=single, caption=Knative~\cite{knative_serving} based on Apache Spark~\cite{knative_serving_java_template}.,  captionpos=b, basicstyle=\ttfamily\tiny, belowskip=-1 \baselineskip, label={fig:knativeexample}]
package com.example.helloworld;
import static spark.Spark.*;

public class HelloWorldApplication {

    public static void main(String args[]) {
        // Change the default port in SparkJava to env. variable port or 8080.
        port(Integer.valueOf(System.getenv().getOrDefault("PORT", "8080")));
        get("/", (req,res) -> "Hello world!");
    }
}



\end{lstlisting}
\end{minipage}  
\vspace{-1mm}
\caption{Comparing \texttt{Java} function handlers across various open-source Serverless Computing Platforms. Handlers are the main code blocks that get executed when a function is triggered.}
\label{sec:javafunctionlistings}
\vspace{-6mm}
\end{figure*}

FaaS provides an attractive cloud model, allowing developers to focus solely on the application logic, while the FaaS platform automatically handles all responsibilities around resource provisioning, scaling, and related management tasks. In addition, the FaaS computing model offers several advantages, such as rapid scalability during request bursts, automatic scaling to zero when resources are unused, and an attractive pricing and development model~\cite{serverlessrise}. However, despite its popularity and several advantages, developers today encounter various challenges while adopting FaaS solutions. These include:

\hspace{-4mm}\circledw{1}.~\textbf{FaaS Platform lock-in}. All current FaaS platforms have their own specific requirements for function development. Particularly, the implementation of functions requires platform-specific function signatures based on the programming language in use. Furthermore, each platform requires unique packages and libraries to handle custom data types and facilitate interaction with its APIs~\cite{Yussupov2019_FaaSPortability}. Listings~\ref{fig:openfaasexample},~\ref{fig:fissionexample},~\ref{fig:nuclioexample}, and~\ref{fig:knativeexample} show the handler implementations for the \texttt{Java} programming language across four different open-source FaaS platforms. All functions are triggered via HTTP requests and return the same response to the user. However, the implemented handlers differ significantly at the code level, making the portability of functions across the different FaaS platforms significantly challenging. As a result, once a FaaS platform has been chosen by the developer, migrating to another platform requires code adaption, reducing developer productivity. 


\hspace{-4mm}\circledw{2}.~\textbf{Diverse deployment configurations}. Individual FaaS platforms exhibit diverse support for configuration parameters during function deployment. These parameters include metadata information such as function name and the deployment namespace, as well as runtime parameters like CPU, memory, and auto-scaling limits. The variation in these parameters across the different FaaS platforms makes the interoperability between them challenging.


\hspace{-4mm}\circledw{3}.~\textbf{Diverse interfaces for FaaS platform interactions}. Individual FaaS platforms have their own unique API interfaces, resulting in distinct interaction requirements. For instance, the process of creating a new function involves calling different API endpoints and transferring different data for each platform. As a result, developers are compelled to grasp the intricacies of each FaaS platform, resulting in diminished productivity and heightened challenges in achieving interoperability.

Enabling portability and interoperability for FaaS functions across different FaaS platforms can easily facilitate use cases such as multi-cloud serverless~\cite{baarzi2021merits} and the seamless deployment and configuration of functions across the edge-cloud continuum~\cite{jindal2021function}. However, existing frameworks in this domain either require code adaptation~\cite{serverless_framework} or support only a specific programming language~\cite{rodrigues2022quickfaas}, impeding their adoption. To this end, our key contributions are:

\begin{itemize}
    \item We present \emph{gFaaS}~\cite{gfaaS}, a novel framework that enables the holistic development, configuration, and management of FaaS functions across different FaaS platforms.
    \item \emph{gFaaS} supports four popular open-source FaaS platforms and enables the development of \emph{generic} platform-independent functions in multiple programming languages. These functions can be seamlessly deployed across any FaaS platform without any modifications.
    \item We demonstrate the viability of our framework by analyzing the overhead of \emph{gFaaS} functions compared to their native platform counterparts for various experiment scenarios.
\end{itemize}
\textbf{Paper Structure}: In \S\ref{sec:gfaaS},
the design goals of \emph{gFaaS}, its system design, and its various components are described. \S\ref{exp:evaluation} presents our evaluation results. In \S\ref{sec:relatedwork}, we present a comparison of \emph{gFaaS} with other frameworks that facilitate portable/inter-operable FaaS functions.
Finally, \S\ref{sec:futurework} concludes the paper and presents an outlook.






\section{gFaaS}
\label{sec:gfaaS}
In this section, we describe our framework \emph{gFaaS} in detail. 

\vspace{-3mm}

\subsection{Design Goals}
\label{sec:designgoals}
To enhance adoption and ease-of-use of \emph{gFaaS}, we chose the following design goals:
\begin{itemize}
    \item Support for multiple FaaS platforms, ensuring ease of extension to seamlessly integrate additional platforms in the future.
    \item Support for multiple programming languages, ensuring easy development and deployment of functions across supported FaaS platforms without modifications.
    \item Support for configuring and managing functions across the different FaaS platforms.
    \item Support for different function invocation triggers such as HTTP and \texttt{gRPC}.
    \item Support for migrating legacy code to \emph{gFaaS}.
\end{itemize}



\vspace{-4mm}

\begin{figure}[t]
\centering
\includegraphics[width=\columnwidth]{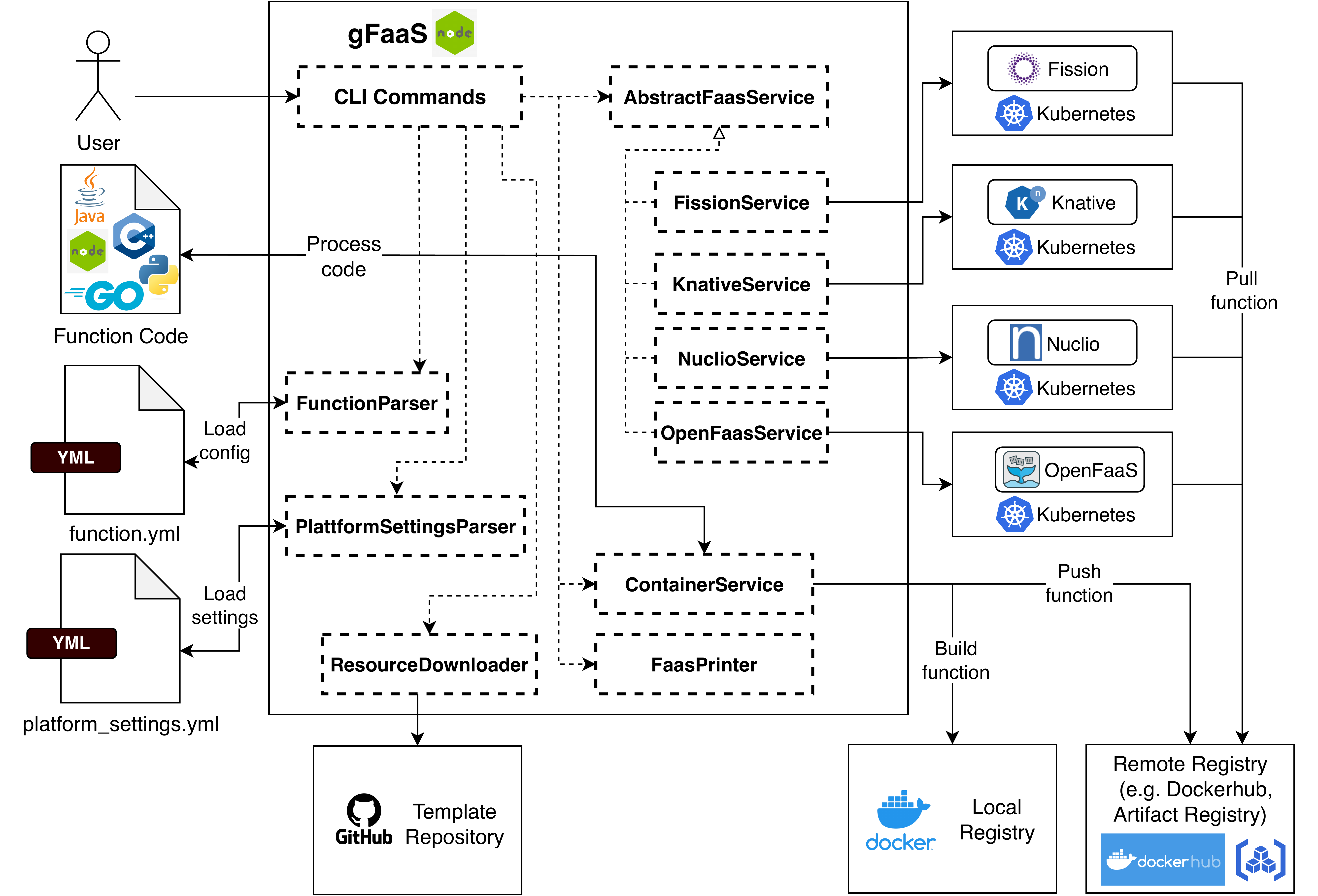}
\vspace{-1mm}
\caption{The different modules within \emph{gFaaS} and their interactions with external components.}
\label{fig:gfaassysdesign}
\shrinkspace
\end{figure}

\subsection{Overview}
\label{sec:imploverview}
\vspace{-1mm}
For the implementation of \emph{gFaaS}, we chose the \texttt{Typescript} programming language for the \texttt{Node.js} runtime environment.  Unlike \texttt{JavaScript}, \texttt{TypeScript} offers the advantage of strong typing, allowing the detection and correction of errors during compilation rather than solely at runtime. Our framework currently supports four open-source FaaS platforms, i.e., OpenFaaS~\cite{openfaas}, Nuclio~\cite{nuclio}, Fission~\cite{fission}, and Knative~\cite{knative_serving}. We chose these platforms due to their popularity and adoption in commercial production environments. For instance, Knative is used in GCF and Google Cloud Run~\cite{google_cloud_run}. Furthermore, \emph{gFaaS} supports the implementation of functions in multiple programming languages, including \texttt{Java}, \texttt{Node.js}, \texttt{Python}, \texttt{Go}, and \texttt{C++}. With \emph{gFaaS}, all function implementations must be containerized into OCI-compliant~\cite{oci} container images. Our framework natively supports building function images using Docker. In addition, our framework provides a CLI interface to easily create, configure, and manage functions across the different platforms.  To easily package and distribute \emph{gFaaS} across multiple operating systems and processor architectures, we have created a container image of \emph{gFaaS}, facilitating its utilization through containerization solutions such as Docker.



\newcommand\YAMLcolonstyle{\color{red}\mdseries}
\newcommand\YAMLkeystyle{\color{black}\bfseries}
\newcommand\YAMLvaluestyle{\color{blue}\mdseries}

\makeatletter


\newcommand\language@yaml{yaml}
\expandafter\expandafter\expandafter\lstdefinelanguage
\expandafter{\language@yaml}
{
frame=single,
captionpos=b,
numbers=left,
framexleftmargin=2mm,
numbersep=1mm,
keywords={true,false,null,y,n},
keywordstyle=\color{darkgray}\bfseries,
basicstyle=\YAMLkeystyle\tiny,                                 
sensitive=false,
comment=[l]{\#},
morecomment=[s]{/*}{*/},
commentstyle=\color{purple}\ttfamily,
stringstyle=\YAMLvaluestyle\ttfamily,
moredelim=[l][\color{orange}]{\&},
moredelim=[l][\color{magenta}]{*},
moredelim=**[il][\YAMLcolonstyle{:}\YAMLvaluestyle]{:},   
morestring=[b]',
morestring=[b]",
literate =    {---}{{\ProcessThreeDashes}}3
{>}{{\textcolor{red}\textgreater}}1
{|}{{\textcolor{red}\textbar}}1
{\ -\ }{{\mdseries\ -\ }}3,
}

\lst@AddToHook{EveryLine}{\ifx\lst@language\language@yaml\YAMLkeystyle\fi}
\makeatother

\newcommand\ProcessThreeDashes{\llap{\color{cyan}\mdseries-{-}-}}

\begin{figure}[t]
\begin{lstlisting}[language=yaml, label={yaml:functionconfig}, caption={Example function \texttt{YAML} configuration file for a \texttt{Java} function with \emph{gFaaS}.}, belowskip=-1.5\baselineskip]
name: gfaas-java
language: java
image: username/gfaas-java:latest
deployRegistry: registry.hub.docker.com
namespace: gfaas-fn
limits:
    memory: 4Gi
    cpu: 2000m
requests:
    memory: 4Gi
    cpu: 2000m
scale:
    knative:
        scaleToZeroDuration: 2m
        min: 1
        max: 5
        
\end{lstlisting}
\vspace{-4mm}
\end{figure}

\vspace{-4mm}

\begin{figure}
\begin{minipage}{0.48\textwidth}
\begin{lstlisting}[language=java, frame=single, caption=\emph{Generic} \texttt{Java} function handler with \emph{gFaaS}.,  captionpos=b, basicstyle=\ttfamily\tiny, belowskip=-1 \baselineskip, label={fig:gFaaSfunction}]
package org.gfaas.core.example;
import org.gfaas.core.model.XFunction;
import org.gfaas.core.model.XRequest;
import org.gfaas.core.model.XResponse;

public class TestFunction extends XFunction {
    @Override
    public XResponse call(XRequest xRequest) {
        var xResponse = new XResponse();
        xResponse.setBody("Hello world!");
        xResponse.setStatusCode(200);
        return xResponse;
    }
}
\end{lstlisting}
\end{minipage}  
\shrinkspace
\vspace{-2mm}
\end{figure}

\subsection{System Design}
\label{sec:sysdesign}

Figure~\ref{fig:gfaassysdesign} presents the different system components within \emph{gFaaS} and their interactions with external components. From the perspective of the user, using our framework is relatively straightforward. A user simply interacts with our framework using the supported CLI commands for all aspects of function handling. In the following, we describe the different components in detail.


\subsubsection{Function Configuration}
\label{sec:funcconfigs}
The function configuration file, written in \texttt{YAML}, enables platform-independent function configuration. It includes important metadata, such as the function name, deployment namespace, CPU and RAM specifications, and scaling parameters. Listing~\ref{yaml:functionconfig} provides an example of a \texttt{Java} function configuration file. The file contains the name of the function's container image (\S\ref{sec:imploverview}) and specifies the image registry where it's stored. Furthermore, it specifies the function's CPU and memory requirements, defined as \emph{requests} and \emph{limits}~\cite{kubernetes_requests_limits}. For instance, the function requires 2 CPU cores and 4 GiB of memory, as shown in Listing~\ref{yaml:functionconfig}. The supported scaling configurations for functions vary across the different FaaS platforms. To address this, \emph{gFaaS} supports platform-specific configuration parameters under the key \texttt{scale} (Line 12). In the provided example, for Knative, it specifies a minimum of one active function instance with a maximum limit of five. Furthermore, after a two-minute idle period, the function instances ($>$1) are terminated and scaled down to zero as specified in the \texttt{scaleToZeroDuration} parameter. If no scaling parameters are specified, then the default configuration values for a platform are selected. The \emph{FunctionParser} subcomponent processes the function configuration file and is also responsible for setting the default parameters.




\subsubsection{Platform Configuration}
\label{sec:platformconfigs}
To configure the different supported FaaS platforms with \emph{gFaaS}, users can provide a platform configuration file written in \texttt{YAML}. This file includes configuration parameters such as the \emph{managementHost} and \emph{managementPort} for accessing the different FaaS platforms and the authentication method. \emph{gFaaS} currently supports three authentication methods: \texttt{basic-auth} using a username and password, \texttt{bearer} with an authentication token, and \texttt{no-auth} enabling unrestricted access to the platforms. The platform configuration file is parsed by the \emph{PlatformSettingsParser} subcomponent as shown in Figure~\ref{fig:gfaassysdesign}.




\subsubsection{Platform-independent function code}
\label{sec:funccode}
To support \emph{generic} functions, \emph{gFaaS} provides uniform classes and interfaces that can be extended to support different FaaS platforms. A generic \emph{gFaaS} function consists of two parts. First, a software package called \emph{gFaaS Core} and second, the function code itself. The \emph{gFaaS Core} contains a preconfigured HTTP server that provides an endpoint for function invocation, along with other essential endpoints required by the different FaaS platforms. For instance, OpenFaaS and Nuclio require liveness and readiness endpoints for querying the function's status. Listing~\ref{fig:gFaaSfunction} shows a \emph{generic} \emph{gFaaS} function in Java. The implemented function does not include any platform-specific dependencies or packages and can be seamlessly deployed on all platforms.




\subsubsection{Supported CLI Commands}
\label{sec:clicommands}
\emph{gFaaS} supports multiple CLI commands for performing platform-independent CRUD operations on functions. These include \texttt{newFunction}, \texttt{build}, \texttt{push}, \texttt{deploy}, \texttt{functions}, \texttt{delete}, and \texttt{invoke}. The descriptions of all supported commands and their functionalities are available in our code repository.




\subsubsection{FaaSService}
\label{sec:faasservice}
The \emph{FaaSService} is an abstract class that provides a uniform interface for supporting the different FaaS platforms. Each supported platform must extend this class and provide implementations for the specified methods. For platform-specific operations, all subcomponents within \emph{gFaaS}, such as the CLI commands (\ref{sec:clicommands}), interact with the different FaaS platforms only via the abstract class functions. This makes internal communication within our framework generic, uniform, and extensible. As a result, support for other FaaS platforms, such as GCF, can be easily integrated into \emph{gFaaS}.


\begin{figure}[t]
\centering
\includegraphics[width=0.8\columnwidth]{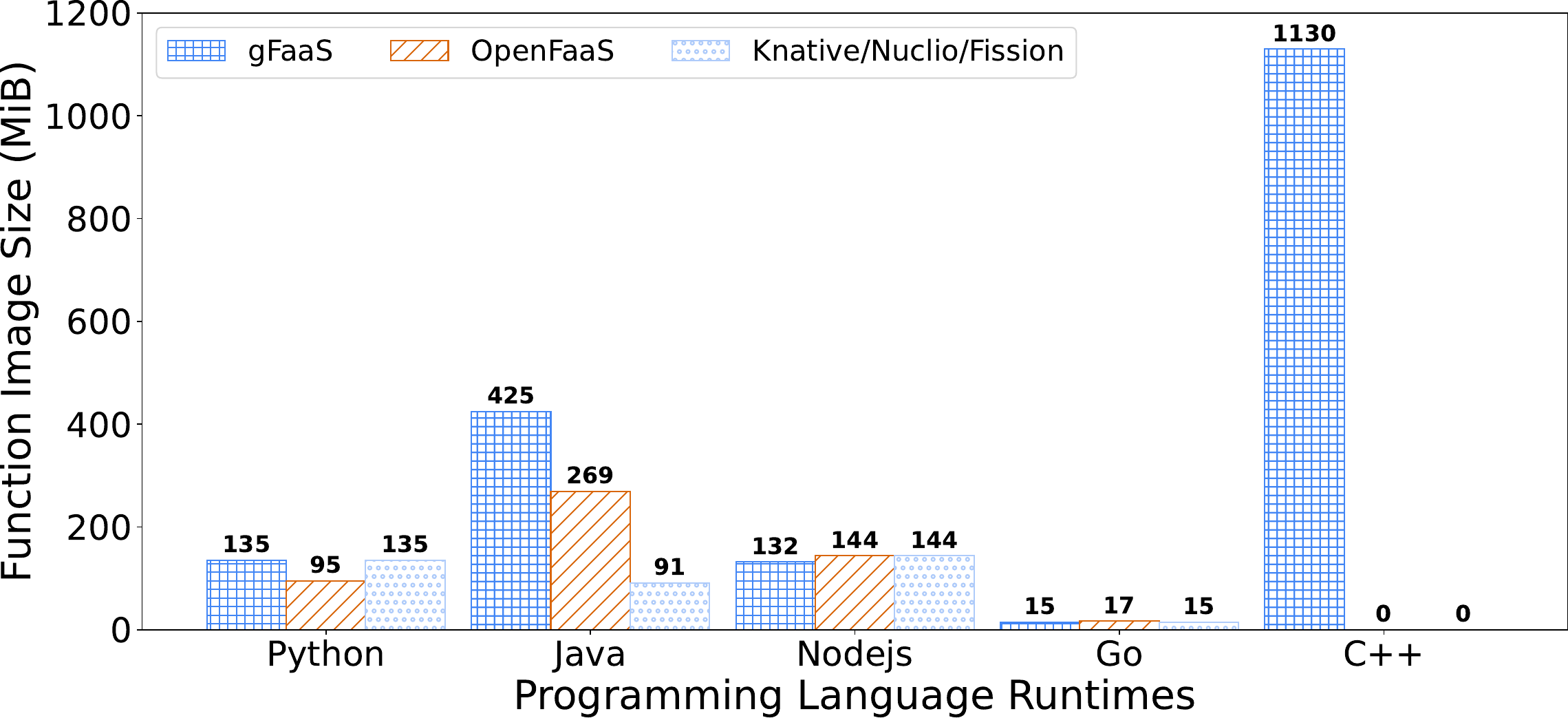}
\vspace{-2mm}
\caption{Comparing function image sizes for the different language runtimes and FaaS platforms.}
\label{fig:fuctionimagesizes}
\shrinkspace
\vspace{-1mm}
\end{figure}

\subsubsection{ContainerService}
\label{sec:containerservice}
The \emph{ContainerService} orchestrates interactions between the function code and both the local and remote container registries. It facilitates the construction of function images and their transfer to a remote registry, while also offering authentication support for these registries. The current implementation of \emph{gFaaS} supports Docker using the \texttt{docker-node} library~\cite{docker_node_library} for building and pushing function images.


\begin{figure*}[t]

 \begin{subfigure}{0.24\textwidth}
    \centering
        \includegraphics[width=\columnwidth]{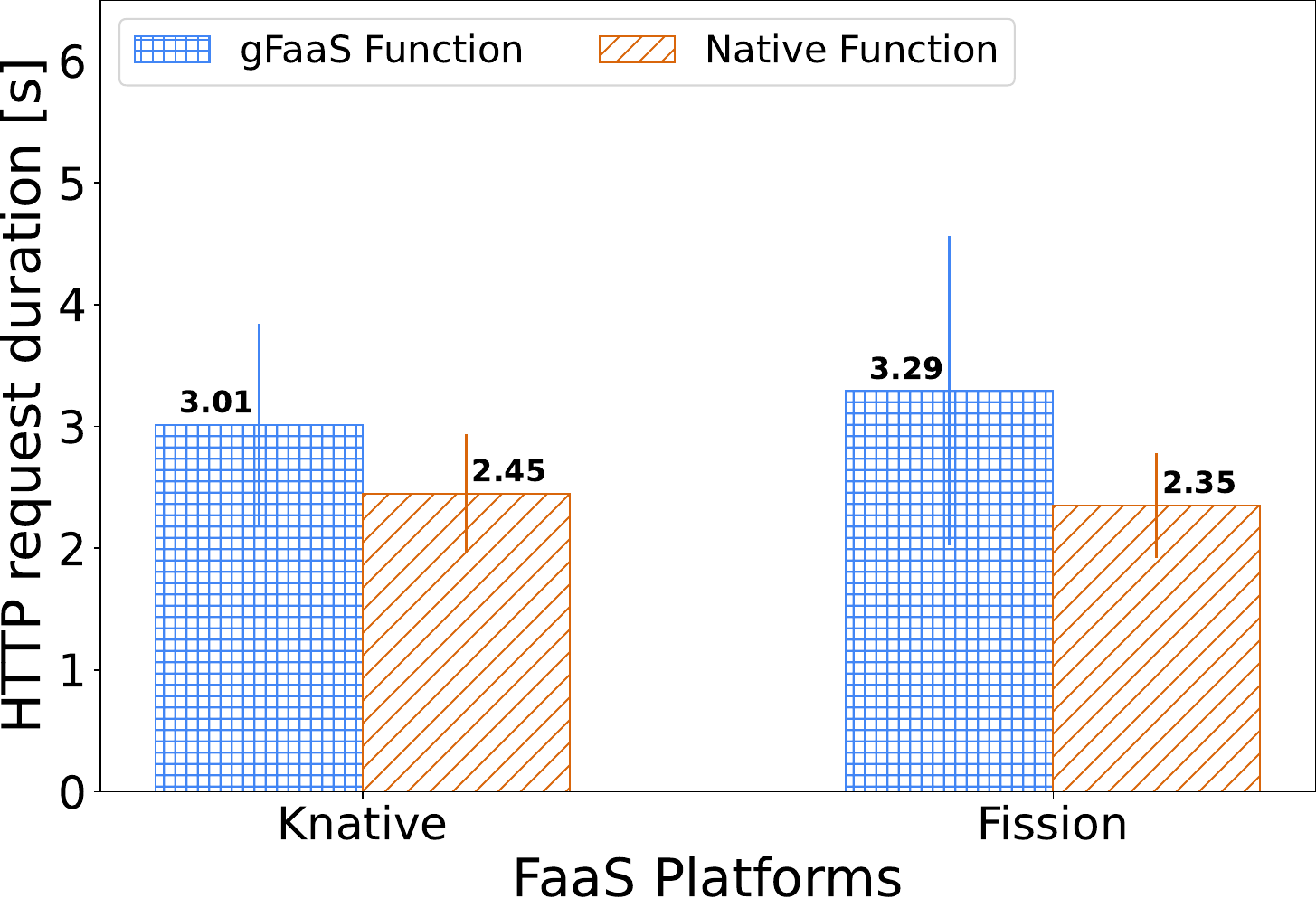}
        \caption{\texttt{Node.js}.}
        \label{fig:nodejscoldstart}
\end{subfigure}
\begin{subfigure}{0.24\textwidth}
    \centering
        \includegraphics[width=\columnwidth]{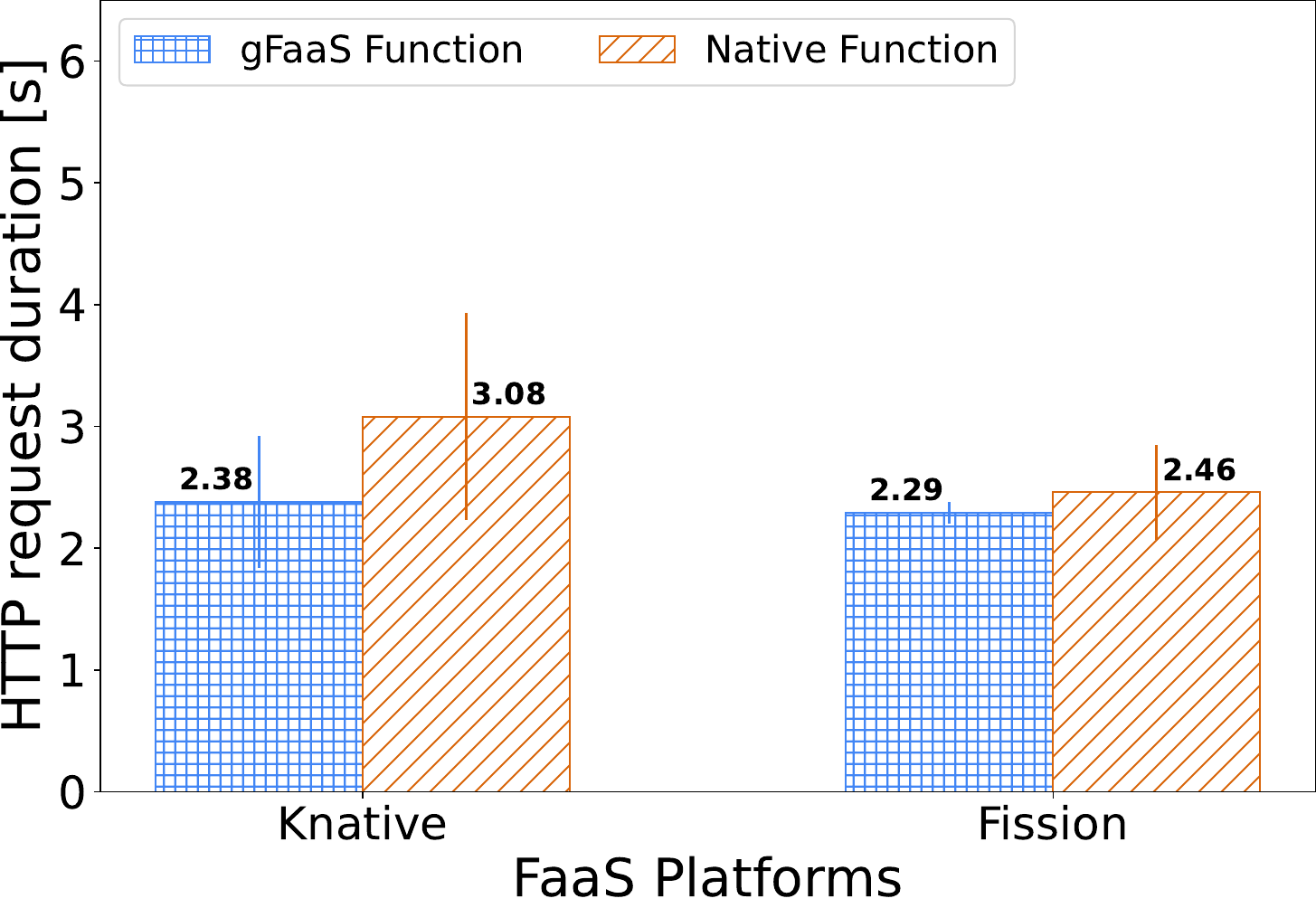}
        \caption{\texttt{Java}.}
        \label{fig:javacoldstart}
\end{subfigure}
 \begin{subfigure}{0.24\textwidth}
    \centering
        \includegraphics[width=\columnwidth]{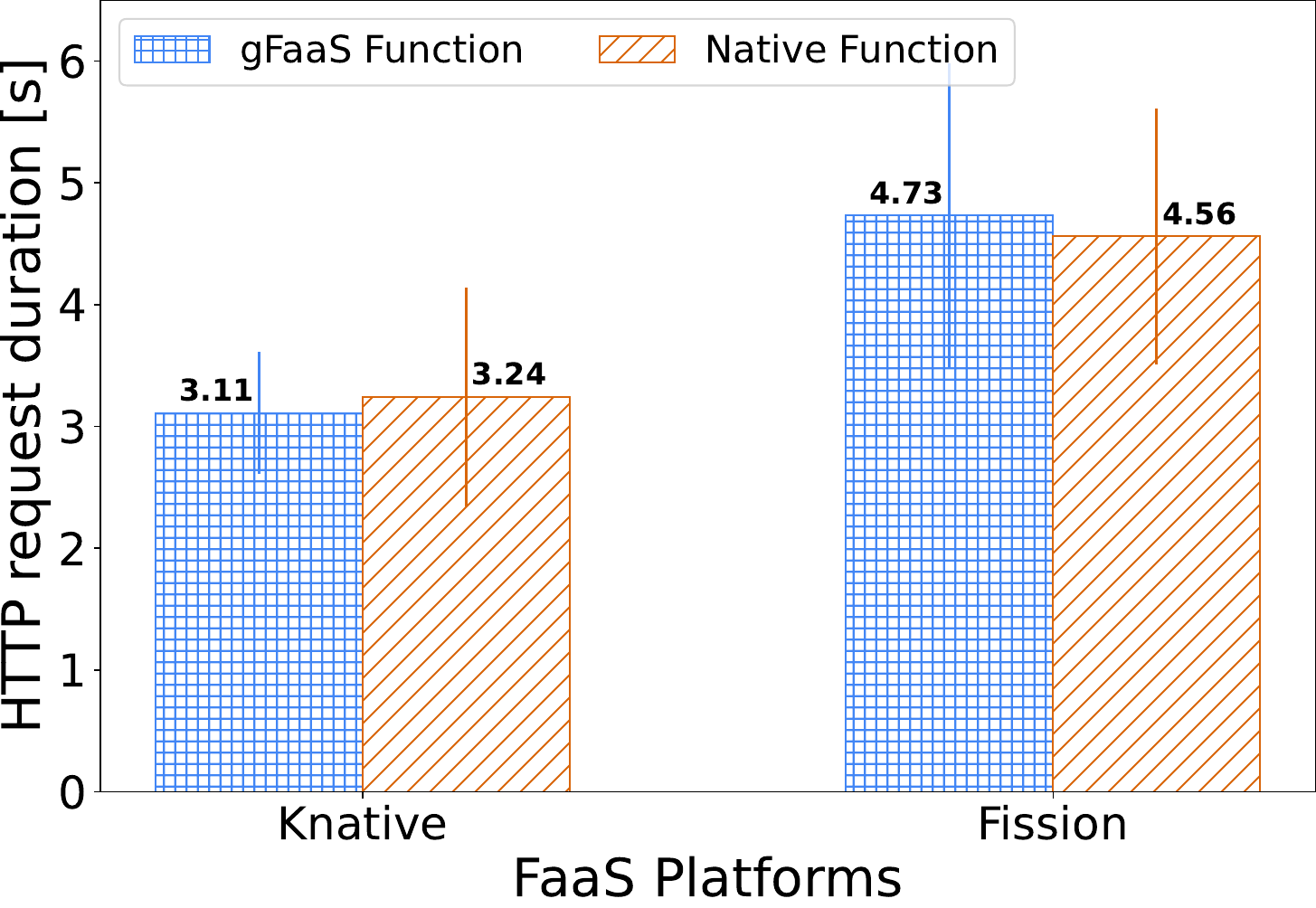}
        \caption{\texttt{Python}.}
        \label{fig:pythoncoldstart}
\end{subfigure}
 \begin{subfigure}{0.24\textwidth}
    \centering
        \includegraphics[width=\columnwidth]{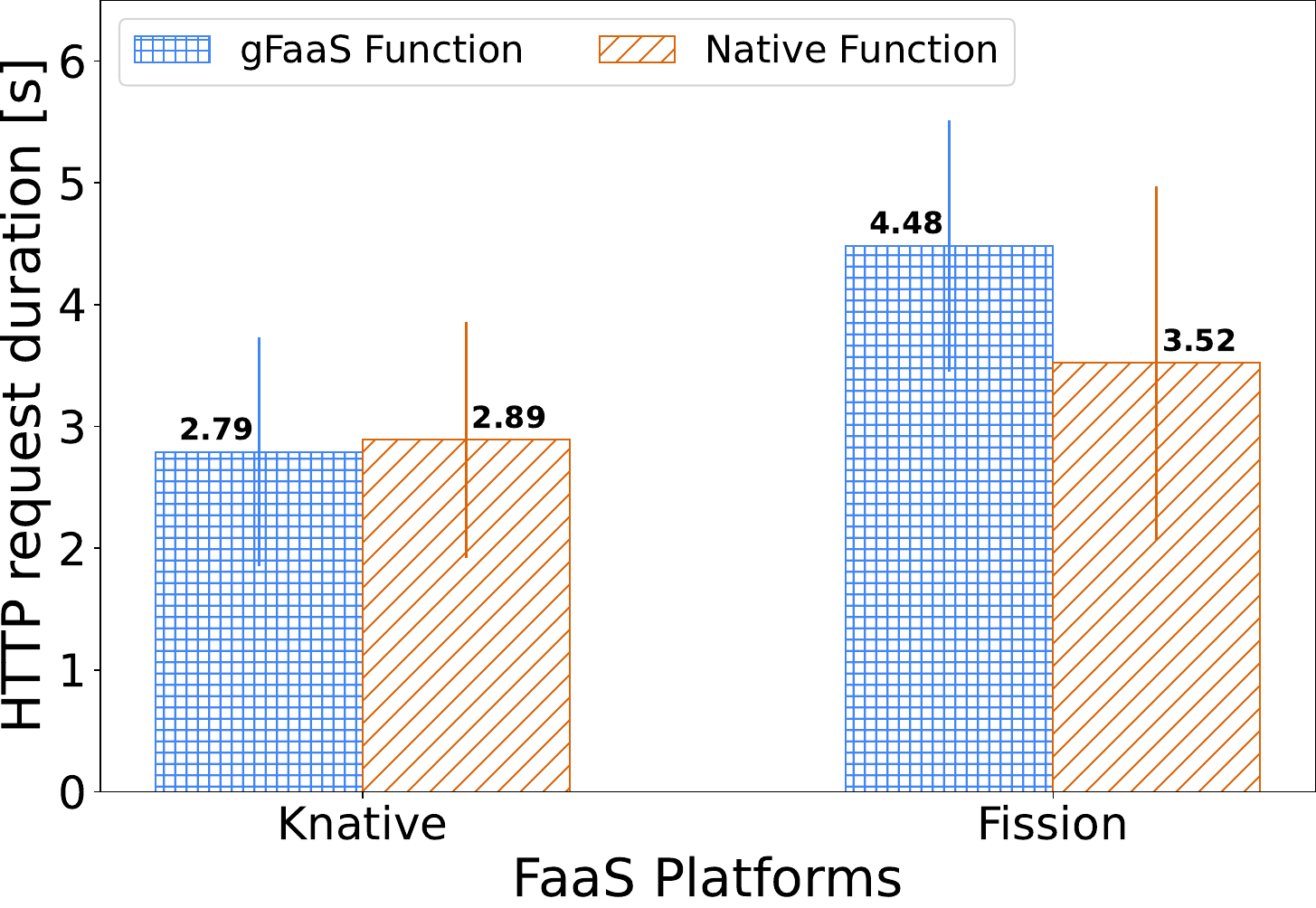}
        \caption{\texttt{Go}.}
        \label{fig:gocoldstart}
\end{subfigure}
\vspace{-2mm}
\caption{Comparing cold start response times for the different language runtimes and FaaS platforms.}
\label{fig:coldstarttimes} 
\shrinkspace
\vspace{-2mm}
\end{figure*}

\begin{figure}[t]
\centering
 \begin{subfigure}{0.24\textwidth}
    \centering
        \includegraphics[width=\columnwidth]{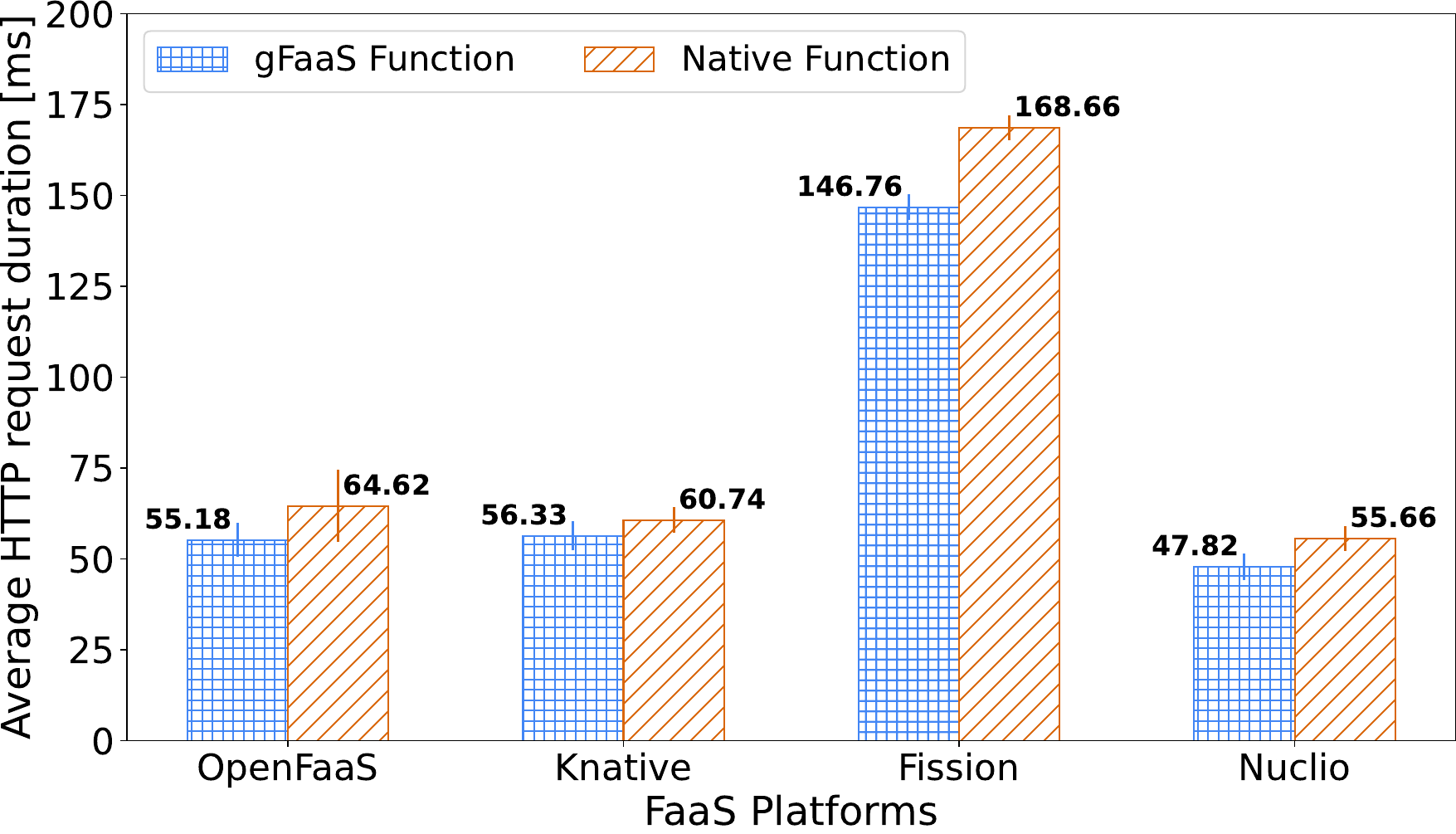}
        \caption{\texttt{Node.js}.}
        \label{fig:pythoncoldstart}
\end{subfigure}
 \begin{subfigure}{0.24\textwidth}
    \centering
        \includegraphics[width=\columnwidth]{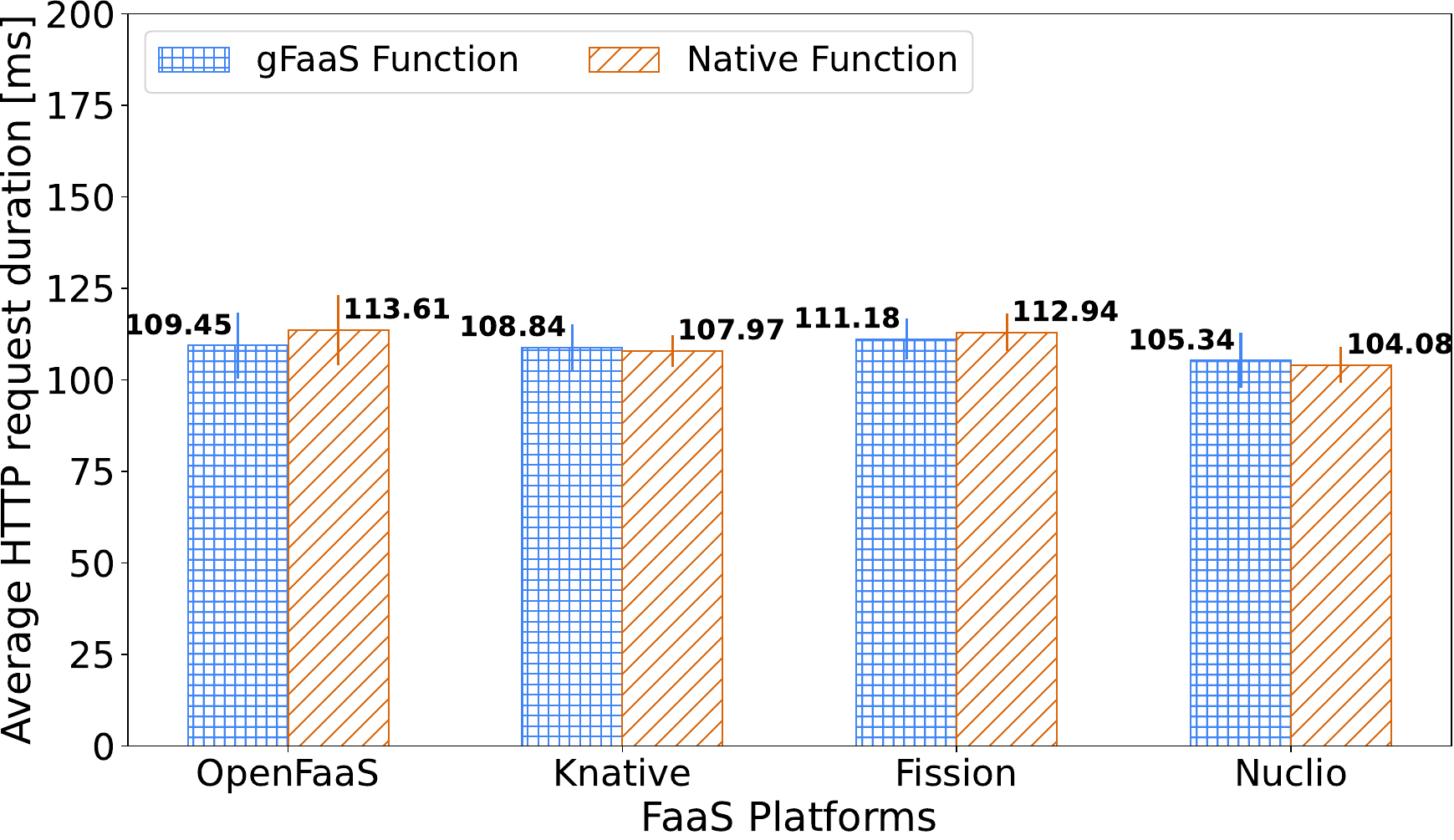}
        \caption{\texttt{Go}.}
        \label{fig:gocoldstart}
\end{subfigure}

\vspace{-2mm}
\caption{Comparing hot start response times for the different language runtimes and FaaS platforms.}
\label{fig:hotstarttimes}
\shrinkspace
\vspace{-2mm}
\end{figure}

\subsubsection{ResourceDownloader}
\label{sec:faasressourcedownloader}

This subcomponent is responsible for interacting with a GitHub repository~\cite{gfaaS_templates} that contains code and templates for creating new functions and migrating legacy code. As a result, this subcomponent is primarily used by the CLI commands \texttt{newFunction} and \texttt{adapt}.



\vspace{-4mm}

\subsection{Supporting legacy code migration}
\label{sec:legacycodemigration}
\vspace{-1mm}
To support the migration of legacy function code to \emph{gFaaS}, our framework uses the \emph{adapter} pattern. The adapter pattern is a widely used technique that allows the integration of legacy code without requiring any modifications. In addition, it offers users complete control over the adaptation process. \emph{gFaaS} provides the \texttt{adapt} CLI command, which users can run in any project of a supported programming language. This command generates and downloads all files required for code migration, along with a detailed \texttt{Readme.md} file that contains instructions for the migration process. This functionality enables existing users of various FaaS platforms, like Knative, to seamlessly utilize \emph{gFaaS} without necessitating modifications to their current function code.




\vspace{-4mm}

\subsection{Supporting \texttt{gRPC}}
\label{sec:supportgrpc}
\texttt{gRPC} is a high-performance remote procedure call framework that enables communication between different services. It relies on \texttt{HTTP/2} and uses protocol buffers to serialize structured data. Our framework currently supports the invocation of functions via \texttt{gRPC} requests only for Knative. This is because the other FaaS platforms do not currently support \texttt{gRPC}. To implement functions with \texttt{gRPC} support, users can leverage the \texttt{newFunction} CLI command, which automatically provisions function templates and essential \texttt{protobuf} files. Additionally, for function deployment, users need to set the \texttt{is\_gRPC} key as 'true' in the function configuration file.






\section{Experimental Evaluation}
\label{exp:evaluation}
\vspace{-1mm}
In this section, we evaluate the overhead of our framework in contrast to utilizing native FaaS platforms across multiple parameters, including image size (\S\ref{sec:imagesizes}), cold start performance (\S\ref{sec:coldstart}), and hot start performance (\S\ref{sec:hotstart}).



\vspace{-4mm}

\subsection{Experimental Setup}
\label{sec:expsetup}
For all our experiments, we use virtual machines (VMs) hosted on the LRZ compute cloud. Each individual FaaS platform is deployed with Kubernetes on a cluster of three VMs. Each VM is configured with 10vCPUs and 45GiB of RAM. In all our experiments, we use a simple function that returns \emph{'Hello World'} as the response to the user. For implementing platform-specific functions for OpenFaaS and Knative, we utilize the templates available in their respective code repositories. However, we faced issues during the build and execution of functions in Nuclio and Fission using their provided templates. As a result, we adapted the Knative function template for these platforms. We repeat all our experiments five times and follow best practices while reporting results.



\vspace{-4mm}

\subsection{Comparing Image Sizes}
\label{sec:imagesizes}
The function image size is not an indicator of a function's performance, 
yet it remains a crucial factor for estimating memory needs and the associated storage costs. Figure~\ref{fig:fuctionimagesizes} shows the function image sizes for the different FaaS platforms and programming languages. For calculating the function image size, we used the tool \texttt{dive}~\cite{dive}. We observe that function images for \texttt{Python}, \texttt{Node.js}, and \texttt{Go} with \emph{gFaaS} are similar in size as compared to the other FaaS platforms. On the other hand, for \texttt{Java}, we observe that the \emph{gFaaS} function image is 58\% larger than the OpenFaaS image and 367\% larger than the images for Knative, Fission, and Nuclio. This can be attributed to the use of different base images for building the function image. OpenFaaS employs \texttt{openjdk:11-jre-slim} (223MB), Knative uses \texttt{openjdk:8-jre-alpine} (85MB), while gFaaS adopts a newer version, \texttt{openjdk:19-jdk-slim} (425MB), which notably represents the smallest available image size for JDK-19. We chose this version since it offers developers access to more features for developing their functions. We omit results for \texttt{C++} for the other FaaS platforms in Figure~\ref{fig:fuctionimagesizes} due to the absence of function templates supporting this language in those platforms.





\vspace{-4mm}

\subsection{Comparing Cold Start performance}
\label{sec:coldstart}
Cold start refers to the duration required to initialize a new 
execution environment for the function upon invocation. This process involves tasks such as downloading the container image, loading the function code, network initialization, and resource allocation. Figure~\ref{fig:coldstarttimes} shows the end-to-end function response times during cold starts for the different programming languages and FaaS platforms. We omit the results for OpenFaaS and Nuclio since they do not support scaling to zero in their open-source community edition implementations. We observe that \emph{gFaaS} functions exhibit negligible overhead during cold starts and perform comparably to native functions across the two FaaS platforms.


\begin{table*}[t]
\begin{adjustbox}{width=2\columnwidth,  center}
\begin{tabular}{|c|c|c|c|}
\hline
\textbf{Attribute}                      & \textbf{gFaaS} (This work)                                                               & \textbf{Serverless}~\cite{serverless_framework}                                                                                                                                                                                                & \textbf{QuickFaaS}~\cite{rodrigues2022quickfaas}                                                                 \\ \hline
Type                                    & CLI, Framework                                                                & CLI, Framework,SaaS                                                                                                                                                                                                & Desktop Application                                                                \\ \hline
Implementation Language                 & \texttt{TypeScript}                                                                        & \texttt{Nodejs}, \texttt{Javascript}                                                                                                                                                                                               & \texttt{Kotlin}, \texttt{Java}                                                                       \\ \hline
Installation and Usage                  & \texttt{npm}, docker                                                                   & \texttt{npm}, Managed cloud                                                                                                                                                                                                 & Install desktop application                                                        \\ \hline
Supported FaaS Platforms                & \begin{tabular}[c]{@{}c@{}}Knative, OpenFaaS, \\ Fission, Nuclio\end{tabular} & \begin{tabular}[c]{@{}c@{}}AWS Lambda, Azure Functions, Tencent Cloud Functions,\\ Google Cloud Functions, Knative, Alibaba Cloud Functions, \\ Cloudflare workers, Fn, Kubeless, OpenWhisk, Spotinst\end{tabular} & \begin{tabular}[c]{@{}c@{}}Google Cloud Functions, \\ Azure Functions\end{tabular} \\ \hline
Supported Language Runtimes             & \texttt{Node.js}, \texttt{Java}, \texttt{Go}, \texttt{C++}, \texttt{Python}                                                 & \texttt{Node.js}, \texttt{Go}, \texttt{Java}, \texttt{PHP}, \texttt{Python}, \texttt{Ruby}, \texttt{Swift}, \texttt{C\#}                                                                                                                                                                    & \texttt{Java}                                                                               \\ \hline
FaaS platform independent function code & \faThumbsOUp                                                                           & \faThumbsDown                                                                                                                                                                                                                 & \faThumbsOUp                                                                                \\ \hline
gRPC Support                            & \faThumbsOUp                                                                           & \faThumbsDown                                                                                                                                                                                                                 & \faThumbsDown                                                                                 \\ \hline
Support for FaaS platform configuration & \faThumbsOUp                                                                           & \faThumbsOUp                                                                                                                                                                                                                & \faThumbsOUp                                                                                \\ \hline
Support for Function configuration      & \faThumbsOUp                                                                           & \faThumbsOUp                                                                                                                                                                                                                & \faThumbsOUp                                                                                \\ \hline
Support for managing functions          & \faThumbsOUp                                                                           & \faThumbsOUp                                                                                                                                                                                                                & \faThumbsOUp                                                                                \\ \hline
Legacy function migration         & \faThumbsOUp                                                                           & \faThumbsDown                                                                                                                                                                                                                 & \faThumbsDown                                                                                 \\ \hline
\end{tabular}
\end{adjustbox}
\caption{Comparing \emph{gFaaS} with other frameworks/tools for implementing and managing serverless functions. \faThumbsOUp Supported. \faThumbsDown No support.}
\label{tab:compother}
\shrinkspace
\end{table*}

\vspace{-4mm}

\subsection{Comparing Hot Start performance}
\label{sec:hotstart}
Hot start occurs when a function instance is already active upon function invocation. To evaluate hot start performance for the different functions and platforms, we use \texttt{k6}~\cite{perftesting}. \texttt{k6} is a developer-centric open-source load and performance regression testing tool. As part of each \texttt{k6} test, two additional parameters are configured, i.e., Virtual Users (VUs), and duration. VUs are the entities in \texttt{k6} that execute the test and make HTTP(s) requests. In our experiments, we use $100$ VUs and run an experiment for five minutes. Figure~\ref{fig:hotstarttimes} shows the hot start response times for \texttt{Node.js} and \texttt{Go} runtimes across the different FaaS platforms. We observe that \emph{gFaaS} functions exhibit minimal overhead during hot starts and perform comparably to native functions across the different FaaS platforms. Due to space limitations, we omit results for  \texttt{Java} and \texttt{Python} programming languages but observe similar performance results.





\section{Related Work}
\label{sec:relatedwork}
\vspace{-1mm}

To facilitate the portability and interoperability of FaaS functions across various platforms, two frameworks/tools exist. These include  \emph{Serverless}~\cite{serverless_framework} and \emph{QuickFaaS}~\cite{rodrigues2022quickfaas}. Table~\ref{tab:compother} presents a comparative analysis between \emph{gFaaS} and these frameworks. 
\emph{gFaaS} offers a CLI for unified function management across various FaaS platforms and facilitates the development of platform-independent functions through templates and packages. The \emph{Serverless} framework also provides a CLI, a framework, and a web interface for function management and development. In contrast, \emph{QuickFaaS} is a desktop application that provides a GUI for creating and managing serverless functions. \emph{gFaaS} and \emph{Serverless} can both be installed via \texttt{npm}, while \emph{QuickFaaS} requires installation as a desktop application with a specific JRE environment. In addition, \emph{gFaaS} can be executed using Docker, enabling high portability across various platforms. 

The \emph{Serverless} framework supports various commercial and open-source FaaS platforms, while \emph{gFaaS} currently supports four popular open-source FaaS platforms. In contrast, \emph{QuickFaaS} only supports GCF and Azure functions. Both \emph{gFaaS} and \emph{Serverless} support the development of functions in various programming languages, while \emph{QuickFaaS} only supports function development in Java. Moreover, \emph{gFaaS} and \emph{QuickFaaS} support the development of platform-independent functions, while for \emph{Serverless}, code modifications are required for executing functions seamlessly across different FaaS platforms. Currently, \emph{gFaaS} is the only framework that supports the development and invocation of \texttt{gRPC}-based functions (\S\ref{sec:supportgrpc}). Furthermore, our framework provides support for migrating legacy code to \emph{gFaaS} using adapters (\S\ref{sec:legacycodemigration}), a capability lacking in both \emph{Serverless} and \emph{QuickFaaS}.

\section{Conclusion \& Future Work}
\label{sec:futurework}
\vspace{-1mm}
In this paper, we presented \emph{gFaaS}, a novel framework that facilitates the uniform development, configuration, and management of functions across diverse FaaS platforms. Our framework supports multiple FaaS platforms and programming languages for function development. Furthermore, it offers a simple CLI interface that can be used by developers for all aspects of function management. In the future, we plan to extend our framework to support commercial FaaS platforms and integrate new invocation triggers, like the addition of items to object storage.







\section{Acknowledgement}
The research leading to these results was funded by the German Federal Ministry of Education and Research (BMBF) in the scope of the Software Campus program under the grant agreement 01IS17049. 




\bibliographystyle{IEEEtran}
\thispagestyle{empty}
\bibliography{parallelpgm}


\end{document}